\documentclass[a4paper,12pt]{article}
\usepackage{amssymb}
\usepackage{graphics}
\usepackage{amsmath}
\usepackage{amsfonts}
\usepackage[utf8]{inputenc}
\usepackage{fullpage}
\usepackage{boxedminipage}
\usepackage{listings}
\usepackage{minitoc}
\usepackage[pdftex]{graphicx}
\usepackage{graphicx}
\def\theequation{\arabic{section}.\arabic{equation}}
\makeatletter \@addtoreset{equation}{section} \makeatother
\renewcommand{\theequation}{\thesection.\arabic{equation}}
\newif\ifpdf \ifx\pdfoutput\undefined \pdffalse
\else \pdfoutput=1 \pdftrue \fi \ifpdf \else \fi

\begin{document}
\ifpdf \DeclareGraphicsExtensions{.pdf, .jpg, .tif} \else %
\DeclareGraphicsExtensions{.eps, .jpg} \fi
\begin{titlepage}

    \thispagestyle{empty}
    \begin{flushright}

        \hfill{\texttt{hep-th/}}
    \end{flushright}

    \vspace{35pt}
    {\LARGE{\textbf{Spin-Bits
and $\mathcal{N}=4$ SYM}} }

        \vspace{60pt}

        {\large{\bf Alessio Marrani}}

        \vspace{30pt}
{~\\\textit{Istituto Nazionale di Fisica Nucleare} (\textit{INFN}),\\  \textit{Laboratori Nazionali di Frascati} (\textit{LNF}), \\
Via Enrico Fermi 40, I-00044 Frascati, Italy}

        \vspace{40pt}

        {ABSTRACT}

    \vspace{10pt}

   We briefly review the spin-bit formalism, describing the
non-planar dynamics of the $\mathcal{N}=4,d=4$ Super Yang-Mills
$SU(N)$ gauge theory.\\After considering its foundations, we apply
such a formalism to the $su(2)$ sector of purely scalar operators.
In particular, we report an algorithmic formulation of a
deplanarizing procedure for local operators in the planar gauge
theory, used to obtain planarly-consistent, testable conjectures for
the higher-loop $su(2)$ spin-bit Hamiltonians.\\Finally, we outlook
some possible developments and applications.
    \vspace{150pt}

    \noindent \textit{Contribution to the Proceedings of the\\43rd Erice International School of Subnuclear Physics\\``Towards New Milestones in our Quest to go Beyond the Standard Model''\\``Ettore Majorana" Foundation and Centre for Scientific Culture,\\Erice, Italy (29 August--7 September 2005)}
\end{titlepage}
\newpage \baselineskip 6 mm

\tableofcontents

\section{Introduction}

Large $N$ physics \cite{'tHooft:1974jz} gained noticeable interest
in the past few years (for a recent review see e.g.
\cite{Tseytlin:2004xa}) due to the AdS/CFT conjecture
enlightenment \cite {Maldacena:1998re,Gubser:1998bc} and, more
recently, to the consideration of various limits of this
correspondence (\cite{Berenstein:2002jq}-\cite
{Bellucci:2004rub}). Initially formulated in the $N\to\infty$
limit, the conjecture in its strong form extends to finite $N$. It
relates the strongly coupled regime of ${\cal N}=4$ SYM to the
weakly coupled string theory and viceversa. This property, which
makes out of this correspondence a very strong and efficient
predictive tool, appeared to be an obstacle in proving the duality
in itself.

Berenstein, Maldacena and Nastase studied in
\cite{Berenstein:2002jq,Berenstein:2002sa} the correspondence in
the neighborood of null geodesics of AdS$_5\times S^5$, where the
geometry appears to resemble that of a gravitational wave
\cite{Blau:2001ne}\nocite{Blau:2002dy}-\cite{Blau:2002mw}. On the
CFT side this corresponds to focusing on SYM operators with a
large ${\cal R}$-charge. The possibility to find a solution of
string theory \cite{Metsaev:2001bj,Metsaev:2002re} in such a
background allows for a quantitative comparison with predictions
coming from perturbative SYM computations
\cite{Kristjansen:2002bb}(see \cite{Spradlin:2003xc} for recent
reviews on the BMN correspondence and references). This led to an
intensive study of the anomalous dimensions of local
gauge-invariant (g.i.) composite operators in $\mathcal{N}=4$,
4-dimensional ($d=4$) Super Yang--Mills (SYM) theory
\cite{Beisert:2003jj}. A real breakthrough was the discovery of
the integrability of the
Hamiltonians governing anomalous dimensions in the planar limit $%
N\rightarrow \infty $ \cite{Minahan:2002ve,Beisert:2004ry}. Then,
these results were extended to $2$ and higher loops
\cite{Serban:2004jf,Beisert:2003jb}. Indeed, the dynamics in the
sector of single-trace bosonic operators of SYM can be mapped into
that of the Heisenberg SO(6) spin one model, so that the matrix of
planar one-loop anomalous dimensions is identified with the spin
chain Hamiltonian \cite{Minahan:2002ve}. The Bethe Ansatz
techniques used for diagonalizing the Hamiltonian become then a
powerful tool in determining anomalous dimensions in the gauge
theory. As it is now clear, there is a one-to-one correspondence
between single-trace operators in SYM theory and spin states in
spin-chain models. The improved understanding of SYM overshadowed,
to some extent, the study of nonplanar contributions. The latter
has to correspond through AdS/CFT to considering the string
production on the AdS side. String bits \cite{Verlinde:2002ig}
were proposed as a model which mimics this feature out of (but not
very far from) the BMN limit. Although the string bit model yields
a good tool for the computation of the relevant bosonic
quantities, it is affected by serious consistency problems related
to the fermionic doubling
\cite{Bellucci:2004rub,Danielsson:2003yc}.

On the SYM side the exact one-loop dilation operator was derived
in \cite{Beisert:2003jj,Minahan:2002ve,Beisert:2003tq}. When
non-planarity is taken into account, single and multi-trace
operators get mixed. This could still not be tested in the string
dual picture.
 Waiting for a better understanding of string physics on
AdS space, one could hope to learn about string interactions there
by exploiting the dual gauge theory picture. This was the main
motivation of the work carried out by the LNF research group over
the past two years. We studied the corresponding spin system which
mixes the integrable spin approach and the string bits one. Such a
theory can be called a \emph{spin bit} model. Since it allows for
dynamical splitting and joining of chains and its variable content
is given by spins, the spin bit model differs from the spin chain
and the string bit ones, although it can be considered as a
mixture of them. In particular, there is no fermion doubling, and
supersymmetry in the spin bit model is consistently implemented.

At $N\to \infty$ the spin Hamiltonian is a local and integrable operator.
 The Hamiltonian and the total spin generator represent the first
 two charges, in the tower of commuting ones, predicted by
integrability \cite{Beisert:2003tq}. Higher charges are given in
terms of higher powers of next-to-nearest spin generators summed
up over the chain. Corrections in ${1\over N}$ spoil locality and
integrability. The Hamiltonian and its higher spin analogs, which
can be interpreted as broken symmetries of the would-be integrable
system, can still be defined in terms of powers of spin
generators. However, now there is no more restriction to
next-to-nearest interactions and the corresponding charges are no
longer commuting among themselves. The role of these broken
charges in the theory near the ``integrable" point $N\to \infty$
remains to be understood.

If the non-planar contributions are considered, the
single-trace sector is not conserved anymore, and one ends up with
trace splitting and joining in the operator mixing
\cite{Beisert:2002ff}. Even in this case one can still consider a
one-to-one map, the so-called spin-bit map, between local g.i.
operators and a spin system
\cite{Bellucci:2004ru,Bellucci:2004qx}. In this case one has to
introduce a set of new degrees of freedom, beyond the spin states,
which describes the linking structure of the sites in the
spin-chain. This can be encoded in a new field, taking values in
the spin-bits permutation group and introducing a new gauge degree
of freedom \cite{Bellucci:2004qx}.\smallskip

In this paper we use conventions and notations of \cite
{Bellucci:2004ru,Bellucci:2004qx,Bellucci:2005ma1}.
\setcounter{equation}0
\section{The $\mathcal{N}=4,d=4$ SYM $SU(N)$ gauge theory}

In the following we will consider a particular quantum field
theory, namely the $\mathcal{N}$=4,$d=4$ SYM $SU(N)$ gauge theory.
Such a theory has the noteworthy property to be
conformally-invariant, due to the vanishing of its beta function
(see e.g. \cite{Lemes:2001}). It has the following field content:
\begin{equation}
\begin{array}{l}
F_{\mu \nu }\text{ gauge field strength, }\mu ,\nu =0,1,2,3; \\
\\
\phi ^{i}\text{ real scalars, }i=1,...,6\text{ (vector repr. of }SO(6)\text{%
);} \\
\\
\lambda _{\alpha }^{\mathcal{A}},\overline{\lambda }_{\mathcal{A}\overset{\cdot }{%
\alpha }}\text{ gauginos, }\mathcal{A}=1,...,\mathcal{N}=4,\text{ \ }\alpha ,%
\overset{\cdot }{\alpha }=1,2.
\end{array}
\end{equation}
All the fields take value in the adjoint representation (repr.) of
the gauge group $SU(N)$,
i.e. for example $\phi ^{i}=\phi _{a}^{i}T^{a}$, where $T^{a}$ ($%
a=1,...,N^{2}-1$) are the generators of $SU(N)$ in the adjoint.
The scalars also span the vector repr. of $SO(6)$, which is the
maximal compact bosonic subgroup of the whole $\mathcal{N}=4$
supergroup $SU(2,2\mid 4)$; moreover, the underlying algebra
$so(6)\sim su(4)$ is the automorphism, or $\mathcal{R} $-symmetry,
algebra of the whole $\mathcal{N}=4,d=4$ superconformal algebra
(SCA) $psu(2,2\mid 4)$. In the following we will use a compact
notation for the $SU(N)$-gauge covariant derivatives of the
fields, namely ($s\in N\cup \left\{ 0\right\} $)
\begin{eqnarray}
\nabla ^{s}\phi &\equiv &\nabla _{\mu _{1}\mu _{2}...\mu _{s}}\phi ^{i},
\notag \\
&&  \notag \\
\nabla ^{s}\lambda &\equiv &\nabla _{\mu _{1}\mu _{2}...\mu _{s}}\lambda
_{\alpha }^{\mathcal{A}},
\end{eqnarray}
and so on. All the elementary fields, as well their derivatives,
can be obtained by acting with generators of the $\mathcal{N}=4$
SCA on the ``primary" fields $\phi ^{i}$.

By adopting a convenient ``philological" nomenclature, we may say that the $%
\mathcal{N}=4,d=4$ SYM ``alphabet" is composed by the set of
``letters''
\begin{equation}
W_{A}\equiv \left\{ \nabla ^{s}F,\nabla ^{s}\phi ,\nabla
^{s}\lambda ,\nabla ^{s}\overline{\lambda }\right\} .
\end{equation}
The components of $W_{A}$ transform in the so-called ``singleton''
(infinite-dimensional) repr. $V_{F}$ of the $\mathcal{N}=4$ SCA.
Out of the ``letters" $W_{A}$ one can build $SU(N)$ g.i.
``words'', i.e. single-trace composite operators given by traces
(in the adjoint of $SU(N)$) of a sequence of ``letters''
$W_{A}$'s. Examples are given by
\begin{equation}
\begin{array}{l}
\mathcal{O}^{i_{1}i_{2}...i_{n}}\equiv Tr\left( \phi ^{i_{1}}\phi
^{i_{2}}...\phi ^{i_{n}}\right) ; \\
\\
\mathcal{O}_{\alpha \overset{\cdot }{\alpha }}^{i_{1}i_{2}...i_{m+1}}\equiv
Tr(\phi ^{i_{1}}\phi ^{i_{2}}...\phi ^{i_{m}}\nabla _{\mu _{1}...\mu
_{n}}\phi ^{i_{m+1}} \\
~~~~~~~~~~~~~~~~~~~~~~~~~\nabla ^{\mu _{1}...\mu _{n-2}}\lambda _{\alpha }^{%
\mathcal{A}}\nabla ^{\mu _{n-1}\mu _{n}}\overline{\lambda }_{\mathcal{A}%
\overset{\cdot }{\alpha }})
\end{array}
\end{equation}
with
\begin{equation}
\nabla ^{\mu }=\eta ^{\mu \nu }\nabla _{\nu },
\end{equation}
where $\eta ^{\mu \nu }$ is the 4-dim. Minkowski metric. Moreover,
out of ``words'' one can produce ``sentences'', which are
sequences of ``words'', i.e. products of single-trace composite
operators, given by products of traces (in the adjoint of $SU(N)$)
of sequences of ``letters''. Some examples are
\begin{eqnarray}
&&
\begin{array}{l}
\mathcal{O}^{i_{1}i_{2}...i_{n_{1}}j_{1}j_{2}...j_{n_{2}}}\equiv Tr\left(
\phi ^{i_{1}}\phi ^{i_{2}}...\phi ^{i_{n_{1}}}\right) Tr\left( \phi
^{j_{1}}\phi ^{j_{2}}...\phi ^{j_{n_{2}}}\right) ; \\
\\
\mathcal{O}_{\overset{\cdot }{\alpha }\overset{\cdot }{\beta }%
}^{i_{1}...i_{n_{2}}\nu }\equiv Tr\left( \nabla _{\mu _{1}...\mu
_{n_{1}}}\phi ^{i_{1}}...\phi ^{i_{n_{2}}}\right) \cdot  \\
~~~~~~~~~~~~~~~~~~~\cdot Tr\left( \nabla ^{\mu _{1}...\mu _{n_{1}-1}}%
\overline{\lambda }_{\mathcal{A}\overset{\cdot }{\alpha }}\right) Tr\left(
\lambda _{\overset{\cdot }{\beta }}^{\mathcal{A}}F^{\mu _{n_{1}}\nu }\right)
.
\end{array}
\notag \\
&&
\end{eqnarray}
The length of a ``word'' or ``sentence'' in $\mathcal{N}=4$ SYM is
defined as the number of ``letters'' $W_{A}$'s composing the
considered trace structure.

Summarizing, the above introduced ``words'' and ``sentences''
correspond
to (possibly multitrace) $SU(N)$ g.i. polynomial composite operators in $\mathcal{N}%
=4,d=4$ SYM. As we will see further below, the spin-bit map gives
an one-to-one spin description of such (multi)trace structures,
and thus allows one to perturbatively calculate the non-planar
($N<\infty $) anomalous dimensions in (certain sectors of) the
considered conformally-invariant gauge theory.
\setcounter{equation}0
\section{The spin-bit model}

A generic $M$-trace g.i. polynomial composite operator of length
$L$ will have the general form
\begin{eqnarray}
\mathcal{O} &\equiv &Tr\left( W_{A_{1}}...W_{A_{L_{1}}}\right) Tr\left(
W_{A_{L_{1}+1}}...W_{A_{L_{1}+L_{2}}}\right) ...  \notag \\
&&...Tr\left( W_{A_{L-L_{M}+1}}...W_{A_{L}}\right) .
\end{eqnarray}
Let us now consider an element of the permutation group $S_{L}$ (of rank $L!$%
) of $L$ elements, namely
\begin{equation}
\gamma \equiv \left( \gamma _{1}\gamma _{2}...\gamma _{L}\right) :\left(
\begin{array}{cccc}
a_{1} & a_{2} & ... & a_{L} \\
a_{\gamma _{1}} & a_{\gamma _{2}} & ... & a_{\gamma _{L}}
\end{array}
\right) ,
\end{equation}
or equivalently
\begin{equation}
S_{L}\ni \gamma =\left( L_{1}\right) \left( L_{2}\right) ...\left(
L_{M}\right) :\sum_{r=1}^{M}L_{r}=L,  \label{gamma}
\end{equation}
where $S_{L_{r}}\ni \left( L_{r}\right) $ is a cyclic permutation of $L_{r}$
elements ($r=1,...,M$). Actually, Eq. (\ref{gamma}) has a deeper meaning,
because in general $S_{L}$ is split in equivalence classes (labelled by $%
L_{1},L_{2}...L_{k}$ such that $\sum_{r=1}^{k}L_{r}=L$) of
permutations consisting of cycles of respective lengths. By
reducing to (products of the) minimal, non-trivial permutational
``bricks'', that is to (products of the) pair-site permutations
$\sigma _{kl}$, $\left( k,l\right) \in \left\{ 1,...,L\right\}
^{2}$ (which simply exchange the $k$-th and $l$-th elements), the
decomposition expressed by Eq. (\ref{gamma}) leads to (planar and)
non-planar (pair-site) permutational identities, extensively
treated in \cite {Bellucci:2005ma2}.

Thus, by suitably choosing $\gamma $, the operator $\mathcal{O}$
may be rewritten as
\begin{equation}
\mathcal{O}=\left( W_{A_{1}}\right) ^{a_{1}a_{\gamma _{1}}}\left(
W_{A_{2}}\right) ^{a_{2}a_{\gamma _{2}}}...\left( W_{A_{L}}\right)
^{a_{L}a_{\gamma _{L}}},
\end{equation}
where the matrix structure (with values in the adjoint repr. of
$SU(N)$) of the ``letters'' $W_{A}$'s is manifest (by convention
the first upper index is a row index, whereas the second is a
column one).

As it may be easily seen, the fundamental features we have to take
into account are the length $L$ of the operator, the number $M$ of
traces, and the ``linking'' configuration expressed by $\gamma $.
By thinking each ``letter'' as sitting on a distinct spin-chain
site, we may therefore write the following equivalence relation
(which we will extensively comment in the following)
\begin{equation}
\mathcal{O}\equiv \left| A_{1},...,A_{L};\gamma \right\rangle ,
\label{equiv1}
\end{equation}
where $A_{k}$ is the direction in $V_{F}$ determined by the ``letter'' $%
W_{A_{k}}$, i.e. the direction of the spin state at the $k$-th
spin-chain site. The r.h.s. of Eq. (\ref{equiv1}) represents a
state in the Hilbert space of a spin-chain model, but with an
explicit extra degree of freedom (represented by $\gamma $),
properly describing the structure of the interconnections among
spin-chain sites: such an ``improved'' spin-chain model will be
called ``spin-bit'' model. Due to the separating action of the
semicolon in Eq. (\ref{equiv1}), the spin-bit Hilbert space $\mathcal{H}%
_{sb} $ naturally gets divided in an usual spin-part
$\mathcal{H}_{sc}$ (the usual spin-chain Hilbert space) and in a
so-called ``linking'' part. As we will see later, this latter will
allow one to correctly take into account
also the non-planar contributions to anomalous dimensions. Eq. (\ref{equiv1}%
) defines the so-called spin-bit map in $\mathcal{N}=4,d=4$ SYM,
whose isomorphicity we are going to discuss.
\setcounter{equation}0
\def\theequation{3.1.\arabic{equation}}
\subsection{Spin part of $\mathcal{H}_{sb}$: the spin-chain picture}

As previously mentioned, the spin part of the ``improved''
spin-chain state, i.e. of the spin-bit state, is given by
\begin{equation}
\mathcal{H}_{sc}\ni \left| A_{1},...,A_{L}\right\rangle =\left|
S_{1},...,S_{L}\right\rangle =\left| S_{1}\right\rangle \otimes ...\otimes
\left| S_{L}\right\rangle .  \label{spin-part-1}
\end{equation}
As it is well known, the spin-chain Hilbert space $\mathcal{H}_{sc}$ is the
tensor product of the one-spin (or, equivalently, one-site) Hilbert spaces $%
\mathcal{H}_{k}$'s. Consequently, as shown by Eq.
(\ref{spin-part-1}), a generic spin-chain state is given by the
tensor product of the one-spin states at each spin-chain site.
$\mathcal{H}_{k}$ is given by the representation space of the
considered symmetry (which in the case at hand will be described
by a subalgebra of $psu(2,2\mid 4)$) at the $k$-th site. Indeed,
$A_{k}\equiv S_{k}$ is the direction in $\mathcal{H}_{k}$
determined by the symmetry of the $\mathcal{N}=4,d=4$ SYM
``letter'' $W_{A_{k}}$ in $%
V_{F}$
\begin{eqnarray}
\mathcal{H}_{sc} &=&\mathcal{H}_{1}\otimes \mathcal{H}_{2}\otimes ...\otimes
\mathcal{H}_{L}=\left( V_{F}\right) _{1}\otimes \left( V_{F}\right)
_{2}\otimes ...\otimes \left( V_{F}\right) _{L},  \notag \\
&&
\end{eqnarray}
where $\left( V_{F}\right) _{k}$ stands for the ``singleton''
representation space of the (relevant subalgebra of the) SCA
$psu(2,2\mid 4)$ at the $k$-th site.

In the following treatment we will assume that the same
representation of the same symmetry algebra will hold at each
spin-chain site. With such an assumption we get
\begin{equation}
\mathcal{H}_{sc}=\left( V_{F}\right) ^{L},
\end{equation}
where $V_{F}$ is the ``singleton'' repr. space of the (relevant
subalgebra of the) SCA $psu(2,2\mid 4)$, which will determine the
symmetry of the spin-chain model.

In $\mathcal{N}=4,d=4$ SYM theory, the most general symmetry is
given by the whole SCA $psu(2,2\mid 4)$, implying that the
dimension of the one-spin Hilbert space $\mathcal{H}_{k}$ is the
same for every spin-chain site, and it is infinite, because $\dim
\left(
V_{F}\right) =\infty $%
\begin{equation}
\dim \left( \mathcal{H}_{sc}\right) =\dim \left( \left( V_{F}\right)
^{L}\right) =\infty ^{L}.
\end{equation}
The situation changes if compact symmetries (having
finite-dimensional unitary representations) are considered (an
example is the $su(2)$ sector which will be extensively treated
later). Also the possibilities to have non-ultralocalizations
and/or different representations of symmetries from site to site
along the chain could be taken into account, but here we will not
deal with such cases (however, see e.g. \cite {Kundu:2002-1}).

Of course, $\mathcal{H}_{k}$ can be endowed with a consistent scalar product
$\left\langle \cdot \mid \cdot \right\rangle $. Consequently, by choosing an
orthonormal basis $\left\{ e_{\alpha }\right\} $ (with $\alpha $ ranging in
a numerable set)
\begin{equation}
\left\langle e_{\alpha }\mid e_{\beta }\right\rangle =\delta _{\alpha \beta
},
\end{equation}
we get
\begin{equation}
S_{k}=S_{k}^{\alpha }e_{\alpha },
\end{equation}
denoting the decomposition of the spin vector $S_{k}$ along the
orthonormal basis $\left\{ e_{\alpha }\right\} $ of the
``singleton'' repr. space $V_{F}$ of the $\mathcal{N}=4$ SCA
$psu(2,2\mid 4)$ at the spin-chain site labelled by $k$.
Therefore, we may rewrite Eq. (\ref{spin-part-1}) as follows:
\begin{eqnarray}
\left| A_{1},...,A_{L}\right\rangle &=&S_{1}^{\alpha _{1}}...S_{L}^{\alpha
_{L}}\left| e_{\alpha _{1}}\right\rangle \otimes ...\otimes \left| e_{\alpha
_{L}}\right\rangle =  \notag \\
&&  \notag \\
&=&\prod_{k=1}^{L}\otimes S_{k}^{\alpha _{k}}\left| e_{\alpha
_{k}}\right\rangle \equiv \left| S\right\rangle \in \mathcal{H}_{sc}.
\label{spin-part-2}
\end{eqnarray}
\setcounter{equation}0
\def\theequation{3.2.\arabic{equation}}
\subsection{The ``linking'' part and the permutational redundance in $%
\mathcal{H}_{sb}$}

On the other hand, by considering only the ``linking'' part of the
spin-bit state, we obtain
\begin{equation}
\left| \gamma \right\rangle \in \zeta _{L},  \label{linking-part-1}
\end{equation}
where $\zeta _{L}$ is nothing but the representation space of the
permutation group $S_{L}$. It may be considered a metrizable space, too, and
consequently it may be endowed with a consistent scalar product $%
\left\langle \gamma \mid \gamma ^{\prime }\right\rangle =\delta _{\gamma
\gamma ^{\prime }}$.

Thus, it would seem reasonable to define the whole Hilbert space
of the
spin-bit model $\mathcal{H}_{sb}$ as the tensor product of the spin part $%
\mathcal{H}_{sc}$ (given by Eq. (\ref{spin-part-2})) and of the
``linking'' part $\zeta _{L}$ (given by Eq.
(\ref{linking-part-1}))
\begin{equation}
\left| A_{1},...,A_{L};\gamma \right\rangle \equiv \left|
A_{1},...,A_{L}\right\rangle \otimes \left| \gamma \right\rangle \in
\mathcal{H}_{sc}\otimes \zeta _{L}.  \label{direct}
\end{equation}
But, by simply doing a tensor product, we would then over-estimate
the spin-bit Hilbert space $\mathcal{H}_{sb}$. Indeed, an extra
symmetry exists,
given by the action of $S_{L}$ on the direct tensor product $\mathcal{H}%
_{sc}\otimes \zeta _{L}$ and determining the following equivalence relation of ``permutational conjugation"%
\footnote{%
Here and below products in $S_{L}$ are understood as
\begin{equation}
\nonumber
\gamma \cdot \sigma \equiv \gamma \sigma =\gamma \cdot \left( \sigma
_{1},...,\sigma _{L}\right) =\left( \sigma _{\gamma _{1}},...,\sigma
_{\gamma _{L}}\right) .
\end{equation}
}
\begin{equation}
\begin{array}{l}
\left| A_{1},...,A_{L};\gamma \right\rangle \sim \left| A_{\sigma
_{1}},...,A_{\sigma _{L}};\sigma \cdot \gamma \cdot \sigma
^{-1}\right\rangle , \\
\\
\forall \left( \gamma ,\sigma \right) \in \left( S_{L}\right) ^{2};
\end{array}
\label{equiv}
\end{equation}
in particular, for $\sigma =\gamma $ we obtain the property of
cyclicity of the trace:
\begin{equation}
\left| A_{1},...,A_{L};\gamma \right\rangle \sim \left| A_{\gamma
_{1}},...,A_{\gamma _{L}};\gamma \right\rangle ,\text{ \ \ }\forall \gamma
\in S_{L}.
\end{equation}

Otherwise speaking, we may define the representation of the action of the
permutational symmetry group $S_{L}$ on the factorized Hilbert space $%
\mathcal{H}_{sc}\otimes \zeta _{L}$ with the operator
\begin{eqnarray}
\widehat{\Sigma }_{\sigma }\left( \left| A_{1},...,A_{L}\right\rangle
\otimes \left| \gamma \right\rangle \right) &\equiv &\left| A_{\sigma
_{1}},...,A_{\sigma _{L}}\right\rangle \otimes \left| \sigma \cdot \gamma
\cdot \sigma ^{-1}\right\rangle .  \notag \\
&&  \label{redund}
\end{eqnarray}
Actually, the action of $\widehat{\Sigma }_{\sigma }$ corresponds
to nothing but an $S_{L}$-covariant relabelling of the spin-chain
site indices. It is a symmetry of the spin-bit model, in the sense
that it can be easily checked that the r.h.s.'s of Eqs.
(\ref{direct}) and (\ref{redund}) describe the same (multi)trace
polynomial composite g.i. operator of length $L$.

It should also be noticed that, due its very definition, the operator $%
\widehat{\Sigma }_{\sigma }$ may be naturally decomposed as the
direct product of two independent operators, acting on distinct
spaces ($\forall \sigma \in S_{L}$)
\begin{equation}
\widehat{\Sigma }_{\sigma }=U_{\sigma }\otimes \widetilde{\Sigma }_{\sigma },
\end{equation}
with the definitions
\begin{equation}
\begin{array}{l}
U_{\sigma }\left| A_{1},...,A_{L}\right\rangle \equiv  \\
\\
\equiv \left( P_{1,\sigma _{1}}\otimes P_{2,\sigma _{2}}\otimes ...\otimes
P_{L,\sigma _{L}}\right) \left| A_{1},...,A_{L}\right\rangle = \\
\\
=\left| A_{\sigma _{1}},...,A_{\sigma _{L}}\right\rangle ,
\end{array}
\label{Usigma}
\end{equation}
where $P_{k,l}\equiv P_{kl}$ is the pair-site index permutation
operator, acting in the spin-chain Hilbert space
$\mathcal{H}_{sc}$ as follows (upperscripts denote the site
positions):
\begin{equation}
\begin{array}{l}
P_{kl}\left| A_{1},...,\overset{k}{A_{k}},...,\overset{l}{A_{l}}%
,...,A_{L}\right\rangle \equiv  \\
\equiv \left| A_{1},...,\overset{l}{A_{k}},...,\overset{k}{A_{l}}%
,...,A_{L}\right\rangle = \\
=\left| A_{1},...,\overset{k}{A_{l}},...,\overset{l}{A_{k}}%
,...,A_{L}\right\rangle ,
\end{array}
\label{Pkl}
\end{equation}
and
\begin{equation}
\widetilde{\Sigma }_{\sigma }\left| \gamma \right\rangle \equiv \left|
\sigma \cdot \gamma \cdot \sigma ^{-1}\right\rangle .  \label{sigmatilde}
\end{equation}

Thus, in order to make the spin-bit map a one-to one (i.e. isomorphic) map,
we have to quotient by this extra symmetry $S_{L}$, obtaining the following
rigorous definition of (state in the) spin-bit Hilbert space:
\begin{equation}
\begin{array}{l}
\mathcal{H}_{sb}\equiv \left\{ \left( V_{F}\right) ^{L}\otimes \zeta
_{L}\right\} /S_{L}\ni \left| A_{1},...,A_{L};\gamma \right\rangle \equiv
\\
\\
\equiv \left\{ \left| A_{1},...,A_{L}\right\rangle \otimes \left| \gamma
\right\rangle \right\} /S_{L}\equiv \left| A_{1},...,A_{L}\right\rangle
\otimes _{S_{L}}\left| \gamma \right\rangle ,
\end{array}
\label{correct-def}
\end{equation}
where $\otimes _{S_{L}}$stands for the direct tensor product, modulo the
action of $S_{L}$ represented by $\widehat{\Sigma }_{\sigma }$.

Therefore, given an arbitrary factorized basis element $\left|
A_{1},...,A_{L}\right\rangle \otimes \otimes \left| \gamma
\right\rangle $, one can find the corresponding element of the
quotient space $\left\{ \left( V_{F}\right) ^{L}\otimes \zeta
_{L}\right\} /S_{L}$, i.e. the corresponding state in the spin-bit
Hilbert space $\mathcal{H}_{sb}$, by ``averaging'' with respect to
the action of $S_{L}$
\begin{equation}
\begin{array}{l}
\left| A_{1},...,A_{L}\right\rangle \otimes _{S_{L}}\left| \gamma
\right\rangle \equiv \left| A_{1},...,A_{L};\gamma \right\rangle \equiv  \\
\\
\equiv \frac{1}{\left| S_{L}\right| }\sum_{\sigma \in S_{L}}\widehat{\Sigma }%
_{\sigma }\left( \left| A_{1},...,A_{L}\right\rangle \otimes \left| \gamma
\right\rangle \right) = \\
\\
=\frac{1}{\left| S_{L}\right| }\sum_{\sigma \in S_{L}}\left( \left|
A_{\sigma _{1}},...,A_{\sigma _{L}}\right\rangle \otimes \left| \sigma \cdot
\gamma \cdot \sigma ^{-1}\right\rangle \right) \equiv  \\
\\
\equiv \widehat{\Pi }\left( \left| A_{1},...,A_{L}\right\rangle \otimes
\left| \gamma \right\rangle \right) ,
\end{array}
\end{equation}
where $\widehat{\Pi }$ is the cyclic symmetry operator, defined as
\begin{equation}
\widehat{\Pi }\equiv \frac{1}{\left| S_{L}\right| }\sum_{\sigma \in S_{L}}%
\widehat{\Sigma }_{\sigma }=\frac{1}{\left| S_{L}\right| }\sum_{\sigma \in
S_{L}}\left( U_{\sigma }\otimes \widetilde{\Sigma }_{\sigma }\right) ,
\label{Pi}
\end{equation}
and $\left| S_{L}\right| =L!$ is the rank of $S_{L}$.

From Eqs. (\ref{Usigma}), (\ref{sigmatilde}) and (\ref{Pi}), it is
not hard to check that $\widehat{\Pi }$ is actually a projective
operator
\begin{equation}
\left( \widehat{\Pi }\right) ^{2}=\widehat{\Pi },
\end{equation}
and that it commutes with permutationally-invariant operators.
Therefore the state $\left| A_{1},...,A_{L};\gamma \right\rangle
$, defined by Eq. (\ref {correct-def}), is $S_{L}$-invariant, as
it has to be in order to correctly estimate the spin-bit Hilbert
space $\mathcal{H}_{sb}$, and therefore to make the spin-bit map
an isomorphic one.

At 1 loop in SYM perturbation theory, it can be also explicitly shown that the ``extra'' symmetry $S_{L}$ of $\mathcal{%
H}_{sb}$ is nothing but a ``gauge" symmetry, in the sense that the
spin-bit model may be seen as arising from the corresponding
spin-chain model by ``gauging'' with respect to the permutational
symmetry $S_{L}$ \cite{Bellucci:2004qx}, where, as previously
mentioned, $L$ is the length of the considered operator, i.e. the
total number of spin-chain sites, and also the total length of the
spin-chain (if unit distance between neighboring sites is
assumed). \setcounter{equation}0
\def\theequation{3.3.\arabic{equation}}
\subsection{Canonical reduction of $S_{L}$}

By recalling Eq. (\ref{gamma}), a generic element $\gamma \in
S_{L}$ may be decomposed (uniquely, up to some possible pair-site
permutational identities \cite{Bellucci:2005ma2}) as follows:
\begin{equation}
S_{L}\ni \gamma =\left( L_{1}\right) \left( L_{2}\right) ...\left(
L_{M}\right) :\sum_{r=1}^{M}L_{r}=L,\text{ }M\leqslant L,
\end{equation}
where $\left( L_{r}\right) $ is a cyclic permutation of $L_{r}$
elements ($r=1,...,M$).
Due to the ``extra'' symmetry $S_{L}$ determining the equivalence relation (%
\ref{equiv}), by choosing $\sigma \in S_{L}$ (in a suitable way,
depending on the starting element $\gamma \in S_{L}$) it is always
possible to reduce $\gamma $ to its canonical form, i.e. to the
form where each spin-chain site index is sent to the immediate
next one modulo cyclicity
\begin{gather}
\gamma =\left( L_{1}\right) \left( L_{2}\right) ...\left( L_{M}\right)
\notag \\
\downarrow  \notag \\
\overset{\left( \sim \right) }{\gamma }=\sigma \cdot \gamma \cdot \sigma
^{-1}=\sigma \left( L_{1}\right) \left( L_{2}\right) ...\left( L_{M}\right)
\sigma ^{-1}:  \notag \\
k_{r}\longmapsto \left[ k_{r}+1\right] \equiv k_{r}+1,\text{ }mod.\text{ }%
L_{r},
\end{gather}
where $k_{r}$ is a spin-chain site index running inside the $r$-th trace.

Even though in what follows we will not restrict ourselves to
consider (only) the canonical form of the permutations, one should
bear in mind that the $S_{L}$-covariant relabelling of site
indices corresponding to the permutational conjugation given by
Eq. (\ref{equiv}) determines an equivalence relation which makes
the switching to canonical permutational forms not implying any
loss of generality. \setcounter{equation}0
\def\theequation{\arabic{section}.\arabic{equation}}
\section{The dilatation operator in $\mathcal{N}=4,d=4$ SYM }

The anomalous dimensions of g.i. operators in the conformally-invariant $%
\mathcal{N}=4,d=4$ SYM gauge theory are given by the action of the
dilatation operator $\Delta $.

In perturbation theory, it may be written as
\begin{equation}
\Delta \left( g_{YM}\right) =\sum_{n=0}^{\infty }H_{2n}\lambda ^{n},
\label{delta-perturb}
\end{equation}
where $g_{YM}$ is the Yang-Mills coupling constant, and
\begin{equation}
\lambda =\lambda \left( g_{YM}\right) \equiv \frac{g_{YM}^{2}N}{16\pi }
\label{'t Hooft}
\end{equation}
is the 't Hooft coupling. $H_{2n}$ is the $n$-loop effective vertex,
determined by an explicit evaluation of the divergencies of the $n$-loop,
2-point function Feynman amplitudes $\left\langle \mathcal{O}(0)\mathcal{O}%
(x)\right\rangle $ in $\mathcal{N}=4,d=4$ SYM.

The first few effective vertices read \cite{Beisert:2003jj}

\begin{equation}
n=0\text{ (tree level): }H_{0}=\Delta _{0A}Tr\left( W_{A}\check{W}%
^{A}\right) ;  \label{n=0}
\end{equation}
\begin{eqnarray}
&&
\begin{array}{l}
n=1\text{ (1-loop level):} \\
\\
H_{2}=-\frac{2}{N}\sum_{j=0}^{\infty }h(j)\left( P_{j}\right) _{CD}^{AB}:Tr%
\left[ W_{A},\check{W}^{C}\right] \left[ W_{B},\check{W}^{D}\right] :,
\end{array}
\notag \\
&&\label{n=1}
\end{eqnarray}
where
\begin{equation}
\left( \check{W}^{A}\right) _{ab}\equiv \frac{\partial }{\partial \left(
W_{A}\right) ^{ba}}
\end{equation}
is the ``letter'' operatorial derivative in $\mathcal{N}=4,d=4$
SYM, such that
\begin{equation}
\left( \check{W}^{A}\right) _{ab}\left( W_{A}\right) ^{bc}=\delta _{a}^{c},
\end{equation}
and $::$ denotes the ``normal-ordering'' of the operators inside,
namely the fact that the derivatives $\left( \check{W}^{A}\right)
_{ab}$ never act on the ``letters'' from the same group inside the
colons. Moreover, $\Delta _{0A}$ stands for the classical (bare)
dimension of the ``letter'' $W_{A}$. For the elementary fields
previously mentioned, it is $\Delta _{0}=1$ for
each scalar field $\phi ^{i}$ and each ($SU(N)$-covariant) derivative, $%
\Delta _{0}=\frac{3}{2}$ for the gauginos and $\Delta _{0}=2$ for the gauge
field strength.

$\left( P_{j}\right) _{CD}^{AB}$ is the (rank 4) $psu(2,2\mid 4)$ projector
to the irreducible module $V_{j}$ in the expansion of the tensor product of
two $\infty $-dim. ``singleton'' $V_{F}$ representations of the $\mathcal{N}%
=4,d=4$ SCA $psu(2,2\mid 4)$
\begin{equation}
V_{F}\otimes V_{F}=\sum_{j=0}^{\infty }V_{j}.
\end{equation}
In general, the first modules $V_{0}$, $V_{1}$and $V_{2}$ contain the
symmetric, antisymmetric and trace components in the tensor product of two
SYM scalars and their superpartners. Higher modules $V_{j}$, $j\geqslant 3$,
contain spin $\left( j-2\right) $ currents and their superpartners. Finally,
$h(j)$ is the $j$-th harmonic number, defined as
\begin{equation}
h(j)\equiv \sum_{s=1}^{j}\frac{1}{s},\text{ }h(0)\equiv0.
\end{equation}
\setcounter{equation}0
\section{The spin-bit Hamiltonian at 1 loop}

In general, by applying the isomorphic spin-bit map to the
dilatation operator of $\mathcal{N}=4,d=4$ SYM $SU(N)$ gauge
theory, we obtain an operator acting on the spin-bit Hilbert space
$\mathcal{H}_{sb}$. Such an operator may be identified with the
spin-bit Hamiltonian; as we will see, it yields a ``deplanarized''
form of the related spin-chain Hamiltonian, in the sense that it
perfectly reproduces the known results from the theory of
spin-chains in the planar limit $N\rightarrow \infty $.

Since on the SYM side the dilatation operator is perturbatively known, we
will correspondingly obtain a perturbatively expanded expression of the
spin-bit Hamiltonian.

At tree level, we trivially get (see Eqs. (\ref{delta-perturb})
and (\ref {n=0}))
\begin{equation}
\Delta _{n=0}=H_{0}=\Delta _{0A}Tr\left( W_{A}\check{W}^{A}\right) ;
\end{equation}
by applying the spin-bit map, i.e. by applying $H_{0}$ on a
generic spin-bit state, we get that the tree-level spin-bit
Hamiltonian $H_{0,sb}$ is simply proportional to the identity,
namely:
\begin{equation}
H_{0,sb}=\Delta _{0}1,
\end{equation}
where, as previously mentioned, $\Delta _{0}$ is the total
classical dimension of the $SU(N)$-g.i. $\mathcal{N}=4$ SYM
(composite) operator uniquely associated to the considered
spin-bit state.

Let us now consider $n=1$, i.e. the 1-loop contribution to $\Delta
$; from Eqs. (\ref{delta-perturb}), (\ref{'t Hooft}) and
(\ref{n=1}) we obtain
\begin{equation}
\begin{array}{l}
\Delta _{n=1}= \\
\\
=-\frac{g_{YM}^{2}}{8\pi ^{2}}\sum_{j=0}^{\infty }h(j)\left(
P_{j}\right) _{CD}^{AB} \\
\\
:Tr\left[ W_{A},\check{W}^{C}\right] \left[
W_{B},\check{W}^{D}\right] :;
\end{array}
\end{equation}
in order to obtain the 1-loop spin-bit Hamiltonian, we simply have
to apply the 1-loop $\mathcal{N}=4$ SYM effective vertex $H_{2}$
(given by Eq. (\ref {n=1})) to a generic spin-bit state
\begin{equation}
H_{2}\left| A_{1},...,A_{L};\gamma \right\rangle .
\end{equation}
In such a way we will map (by means of the spin-bit isomorphic
correspondence) $H_{2}$ to the 1-loop spin-bit Hamiltonian
$H_{2,sb}$. Clearly, since $H_{2}$ is a second-order differential
operator (it contains two operatorial derivatives), the Leibnitz
rule will decompose the result in a sum over all possible couples
of spin-chain sites:
\begin{equation}
\begin{array}{l}
H_{2}\left| A_{1},...,A_{L};\gamma \right\rangle = \\
\\
=-\frac{2}{N}\sum_{j=0}^{\infty }h(j)\left( P_{j}\right) _{CD}^{AB} \\
\\
:Tr\left[ W_{A},\check{W}^{C}\right] \left[ W_{B},\check{W}^{D}\right]
:\left| A_{1},...,A_{L};\gamma \right\rangle = \\
\\
=\sum_{k,l=1}^{L}H_{2,kl}\left| A_{1},...,\overset{k}{A_{k}},...,\overset{l}{%
A_{l}},...,A_{L};\gamma \right\rangle ,
\end{array}
\label{rot}
\end{equation}
where $H_{2,kl}$ is nothing but the restriction of the 1-loop
effective vertex to the couple of sites $\left( k,l\right) \in
\left\{ 1,...,L\right\} ^{2}$, and it will be later identified, by
the spin-bit map, with the two-site 1-loop spin-bit Hamiltonian.

A number of technical, permutational results are used in the
explicit calculations; they respectively read:
\begin{eqnarray}
\text{Fission formula} &:&\text{ \ \ }Tr\left( A\check{W}^{C}BW_{D}\right)
=\delta _{D}^{C}Tr\left( A\right) Tr(B);  \notag \\
&&
\end{eqnarray}
\begin{eqnarray}
\text{Fusion formula} &:&\text{ \ \ }Tr\left( A\check{W}^{C}\right) Tr\left(
W_{D}B\right) =\delta _{D}^{C}Tr\left( AB\right) ,  \notag \\
&&
\end{eqnarray}
where $A$ and $B$ are supposed not to depend on $W$'s. An useful property
(holding true for any permutation $\gamma $ and for any pair-site
permutation $\sigma _{kl}$ in $S_{L}$) is
\begin{equation}
\sigma _{kl}\gamma =\gamma \sigma _{\gamma _{k}\gamma _{l}};
\end{equation}
by using it, Eq. (\ref{equiv}) yields
\begin{equation}
\begin{array}{l}
\left| A_{1},...,\overset{k}{B},...,\overset{l}{A},...,A_{L};\gamma \sigma
_{kl}\right\rangle = \\
\\
=\left| A_{1},...,\overset{k}{A},...,\overset{l}{B},...,A_{L};\gamma \sigma
_{\gamma _{k}\gamma _{l}}\right\rangle
\end{array}
\end{equation}
or, in terms of operators
\begin{equation}
P_{kl}\Sigma _{kl}=\Sigma _{\gamma _{k}\gamma _{l}},
\end{equation}
where $P_{kl}$ is the $\left( k,l\right) $-site permutation
operator acting on $\mathcal{H}_{sc}$ defined by Eq. (\ref{Pkl}),
and $\Sigma _{k,l}\equiv\Sigma _{kl}$ is the
(1-loop) chain ``splitting and joining'' (or ``twist'') operator, acting on $%
\zeta _{L}$, and defined as
\begin{equation}
\Sigma _{kl}\left| \gamma \right\rangle \equiv \left\{
\begin{array}{ll}
\left| \gamma \sigma _{kl}\right\rangle , & k\neq l \\
N\left| \gamma \right\rangle , & k=l
\end{array}
\right. ,
\end{equation}
or equivalently
\begin{equation}
\Sigma _{kl}=N\delta _{kl}+\left( 1-\delta _{kl}\right) \overline{\Sigma }%
_{kl},\text{ with }\overline{\Sigma }_{kl}\left| \gamma \right\rangle
=\left| \gamma \sigma _{kl}\right\rangle .  \label{sigma-decomp}
\end{equation}
The factor $N$ in the case $k=l$ appears because splitting and joining a
trace/chain at the same site leads to a new chain of length zero, whose
corresponding trace is $Tr\left( 1\right) =N$, because $1$ stands for the
identity in the adjoint repr. of the gauge group $SU(N)$.

The final result is the 1-loop spin-bit Hamiltonian
\begin{equation}
H_{2,sb}=\frac{1}{2N}\sum_{\underset{(k\neq l)}{k,l=1}}^{L}H_{kl}\left(
\Sigma _{\gamma _{k}l}+\Sigma _{k\gamma _{l}}-\Sigma _{kl}-\Sigma _{\gamma
_{k}\gamma _{l}}\right)   \label{rot-rot}
\end{equation}
or, using the canonical form of the permutation $\gamma \in
S_{L}$,
\begin{equation}
\begin{array}{l}
H_{2,sb}= \\
\\
=\frac{1}{2N}\sum_{\underset{(k\neq l)}{k,l=1}}^{L}H_{kl}\left( \Sigma _{%
\left[ k+1\right], l}+\Sigma _{k,\left[ l+1\right] }-\Sigma _{k,l}-\Sigma _{%
\left[ k+1\right], \left[ l+1\right] }\right) ,
\end{array}
\label{rot-rot-rot}
\end{equation}
where $H_{k,l}\equiv H_{kl}\equiv H_{kl,sb}$ is the two-site Hamiltonian, acting on $%
\mathcal{H}_{sc}$, and defined as follows ($k\neq l$):
\begin{equation}
\begin{array}{l}
H_{kl}\left| A_{1},...,A_{L}\right\rangle \equiv  \\
\\
\equiv 4\sum_{j=0}^{\infty }h(j)\left( P_{j}\right) _{A_{k}A_{l}}^{AB}\left|
A_{1},...,\overset{k}{A},...,\overset{l}{B},...,A_{L}\right\rangle .
\end{array}
\label{Hkl}
\end{equation}
Notice that $P_{kl}$ and $H_{kl}$ act on $\mathcal{H}_{sc}$, whereas $\Sigma
_{kl}$ acts on $\zeta _{L}$, and therefore
\begin{eqnarray}
\left[ P_{kl},\Sigma _{mn}\right]  &=&0=\left[ H_{kl},\Sigma _{mn}\right] ,%
\text{ \ \ }\forall \left( k,l,m,n\right) \in \left\{ 1,...,L\right\} ^{4}.
\notag \\
&&
\end{eqnarray}
By comparing Eq. (\ref{rot}) with Eqs. (\ref{rot-rot}) and (\ref{rot-rot-rot}%
), and by disregarding the degenerate case of coinciding sites
$k=l$ (this can be shown not implying any loss of generality), we
may conclude that
\begin{equation}
\begin{array}{l}
H_{2}\left| A_{1},...,\overset{k}{A_{k}},...,\overset{l}{A_{l}}%
,...,A_{L};\gamma \right\rangle = \\
\\
=\sum_{\underset{(k\neq l)}{k,l=1}}^{L}H_{2,kl}\left| A_{1},...,\overset{k}{%
A_{k}},...,\overset{l}{A_{l}},...,A_{L};\gamma \right\rangle = \\
\\
=\frac{1}{2N}\sum_{\underset{(k\neq l)}{k,l=1}}^{L}H_{kl}\left( \Sigma
_{\gamma _{k}l}+\Sigma _{k\gamma _{l}}-\Sigma _{kl}-\Sigma _{\gamma
_{k}\gamma _{l}}\right)  \\
\\
\left| A_{1},...,\overset{k}{A_{k}},...,\overset{l}{A_{l}},...,A_{L};\gamma
\right\rangle \Leftrightarrow  \\
\\
\Leftrightarrow H_{2,kl}\left( \gamma \right) \equiv H_{2,kl}= \\
\\
=\frac{1}{2N}H_{kl}\left( \Sigma _{\gamma _{k}l}+\Sigma _{k\gamma
_{l}}-\Sigma _{kl}-\Sigma _{\gamma _{k}\gamma _{l}}\right) ,\text{ } \\
\\
\forall \left( k,l\right) \in \left\{ 1,...,L\right\} ^{2},k\neq l,
\end{array}
\end{equation}
where, as previously mentioned, $H_{2,kl}$ is the ($\gamma
$-dependent)
restriction of the 1-loop spin-bit Hamiltonian to the couple of sites $%
\left( k,l\right) \in \left\{ 1,...,L\right\} ^{2}$, $k\neq l$.
\setcounter{equation}0
\def\theequation{5.1.\arabic{equation}}
\subsection{The planar limit}

From Eq. (\ref{sigma-decomp}), the planar limit $N\rightarrow
\infty $ affects just the (1-loop) ``twist'' operator in the
following way:
\begin{equation}
\lim_{N\rightarrow \infty }\frac{1}{N}\Sigma _{kl}=\delta _{kl}.
\label{sigma-planar}
\end{equation}
Therefore the planar contributions to the 1-loop spin-bit
Hamiltonian come from terms involving $\Sigma _{kk}$, i.e. from
the cases $l=\gamma _{k}$ and $k=\gamma _{l}$ (because $k\neq l$);
the final result is
\begin{equation}
\lim_{N\rightarrow \infty }H_{2,sb}=\sum_{k=1}^{L}H_{k\gamma
_{k}}=\sum_{k=1}^{L}H_{k,\left[ k+1\right] }=H_{2,sc}.
\end{equation}
Otherwise speaking, by construction the planar limit of the 1-loop
spin-bit Hamiltonian coincides with the 1-loop spin-chain
Hamiltonian $H_{2,sc}$. \setcounter{equation}0
\def\theequation{\arabic{section}.\arabic{equation}}
\section{Spin-bits in the $su(2)$ sector of $\mathcal{N}=4$ SYM}

We will now consider the ``minimal'' sector of $\mathcal{N}=4,d=4$
SYM $SU(N) $ gauge theory, made by ``purely-scalar'' g.i.
(polynomial) operators, i.e. by operators generated only by two
holomorphic combinations of the real SYM scalars, which may be
defined as
\begin{equation}
\left\{
\begin{array}{l}
\Phi \equiv \phi ^{5}+i\phi ^{6}, \\
\\
Z\equiv \phi ^{1}+i\phi ^{2}.
\end{array}
\right.
\end{equation}
Thus, $\Phi $ and $Z$ will be the only SYM ``letters'' used to
compose ``words'' and ``sentences'' in such an operator sector,
which may be shown to be closed under the operator mixing due to
perturbative renormalization of the theory. $\Phi $ and $Z$ will
transform in the $2$-dim. $s=1/2$
fundamental repr. of $su(2)$, and therefore the whole sector will be $su(2)$%
-symmetric. Notice that $su(2)$ is the smallest non-trivial bosonic compact
subalgebra of the whole $\mathcal{N}=4$ SCA $psu(2,2\mid 4)$, and the
following chain of inclusions holds:
\begin{eqnarray}
su(2) &\subset &so(6)\sim su(4)\subset so(4,2)\oplus su(4)\subset
psu(2,2\mid 4).  \notag \\
&&
\end{eqnarray}
A generic $M\left( \leqslant L\right) $-trace g.i. operator in
such a $su(2)$ closed subsector reads
\begin{equation}
\mathcal{O}\equiv \underset{L_{1}\text{ ``letters''}}{Tr\left(
\Phi Z\Phi \Phi ...\right) }\underset{L_{2}\text{
``letters''}}{Tr\left( \Phi \Phi \Phi Z...\right)
}...\underset{L_{M}\text{ ``letters''}}{Tr\left( \Phi ZZ\Phi
Z...\right) ,}  \label{OM}
\end{equation}
with $\sum_{r=1}^{M}L_{r}=L$.

As previously shown, we may use the isomorphic spin-bit map to
equivalently represent $\mathcal{O}$ as
\begin{equation}
\begin{array}{l}
\mathcal{O}=\left( \phi ^{i_{1}}\right) ^{a_{1}a_{\gamma _{1}}}\left( \phi
^{i_{2}}\right) ^{a_{2}a_{\gamma _{2}}}...\left( \phi ^{i_{L}}\right)
^{a_{L}a_{\gamma _{L}}}\equiv  \\
\\
\equiv \left| S;\gamma \right\rangle \equiv \left| S\right\rangle \otimes
_{S_{L}}\left| \gamma \right\rangle ,
\end{array}
\end{equation}
where now the range of all ``$i$-indices'' of the scalars is
$\left\{ \widehat{1},\widehat{2}\right\} $, with $\phi
^{\widehat{1}}\equiv \Phi $
and $\phi ^{\widehat{2}}\equiv Z$ by convention. In the case of Eq. (\ref{OM}%
) we have
\begin{equation}
S_{L}\ni \gamma =\left( L_{1}\right) \left( L_{2}\right) ...\left(
L_{M}\right) ,
\end{equation}
with $\left( L_{r}\right) $ denoting a cyclic permutation of $L_{r}$
elements, $r=1,...,M\leqslant L$. The correspondence operated by the
spin-bit map is completed by associating to each spin-chain site the spin
value $\left| -1/2\right\rangle $ if we find $\Phi $ there, and $\left|
1/2\right\rangle $ if we find $Z$.

By specializing the general expression (\ref{Hkl}) of the two-site
Hamiltonian to the case of $2$-dim. $s=1/2$ (representation of)
$su(2)$ symmetry, the final result is simply \cite{Beisert:2003jj}

\begin{equation}
H_{kl,su(2)}=2\left( 1-P_{kl}\right) ,
\end{equation}
and therefore, by substituting it in the general formulae
(\ref{rot-rot}) and (\ref{rot-rot-rot}), we obtain the 1-loop
$su(2)$ spin-bit Hamiltonian $H_{2,sb,su(2)}\equiv H_{2,su(2)}$
\begin{eqnarray}
H_{2,su(2)} &=&\frac{1}{2N}\sum_{k,l=1}^{L}H_{kl,su(2)}\left( \Sigma
_{\gamma _{k}l}+\Sigma _{k\gamma _{l}}-\Sigma _{kl}-\Sigma _{\gamma
_{k}\gamma _{l}}\right) =  \notag \\
&&  \notag \\
&=&\frac{2}{N}\sum_{k,l=1}^{L}\left( 1-P_{kl}\right) \Sigma _{k\gamma _{l}}.
\label{H2,su(2)}
\end{eqnarray}
In the planar limit, by recalling Eq. (\ref{sigma-planar}) we get
\begin{eqnarray}
&&
\begin{array}{l}
\lim_{N\rightarrow \infty }H_{2,su(2)}=2\sum_{k,l=1}^{L}\left(
1-P_{kl}\right) \lim_{N\rightarrow \infty }\frac{1}{N}\Sigma _{k\gamma _{l}}=
\\
\\
=2\sum_{k,l=1}^{L}\left( 1-P_{kl}\right)  \\
\\
\lim_{N\rightarrow \infty }\frac{1}{N}\left[ N\delta _{k\gamma _{l}}+\left(
1-\delta _{k\gamma _{l}}\right) \overline{\Sigma }_{k\gamma _{l}}\right] =
\\
\\
=2\sum_{k=1}^{L}\left( 1-P_{k\gamma _{k}}\right) =\sum_{k=1}^{L}H_{k\gamma
_{k},su(2)}= \\
\\
=H_{2,sc,su(2)}.
\end{array}
\notag \\
&&
\end{eqnarray}
In other words, the planar limit of the 1-loop $su(2)$ spin-bit
Hamiltonian coincides with the integrable $XXX_{s=1/2}$ Heisenberg
$su(2)$ spin-chain Hamiltonian $H_{2,sc,su(2)}$
\cite{Faddeev:1996}.

The expression (\ref{H2,su(2)}) of $H_{2,su(2)}$ might also be
obtained by starting from the known combinatorial formula of the 1-loop $%
\mathcal{N}=4,d=4$ SYM effective vertex on the $su(2)$-symmetric
closed subsector of ``purely scalar'' composite (polynomial) g.i.
operators, reading \cite{Beisert:2003tq}
\begin{equation}
H_{2}=-\frac{4}{N}:Tr\left( \left[ \Phi ,Z\right] \left[ \check{\Phi},\check{%
Z}\right] \right) :,
\end{equation}
where
\begin{equation}
\left\{
\begin{array}{l}
\left( \check{\Phi}\right) ^{ab}\equiv \frac{\partial }{\partial \Phi ^{ba}},
\\
\\
\left( \check{Z}\right) ^{ab}\equiv \frac{\partial }{\partial Z^{ba}}.
\end{array}
\right.
\end{equation}
\setcounter{equation}0
\def\theequation{\arabic{section}.\arabic{equation}}
\section{$su(2)$ spin-bits at 2 loops}

The $su(2)$ sector of $\mathcal{N}=4,d=4$ SYM is the only one, as
far as we know, for which a detailed treatment of non-planar
($N<\infty $) 2-loop anomalous dimensions has been given. This is
due to the noteworthy fact that a combinatorial formula of the
2-loop $\mathcal{N}=4,d=4$ SYM effective vertex on the
$su(2)$-symmetric closed subsector of ``purely scalar'' composite
(polynomial) g.i. operators is known \cite{Beisert:2003tq}:
\begin{eqnarray}
H_{4} &=&\frac{2}{N^{2}}\left[
\begin{array}{l}
:Tr\left( \left[ Z,\Phi \right] \left[ \check{Z},\left[ Z,\left[ \check{Z},%
\check{\Phi}\right] \right] \right] \right) :+ \\
\\
+:Tr\left( \left[ Z,\Phi \right] \left[ \check{\Phi},\left[ \Phi ,\left[
\check{Z},\check{\Phi}\right] \right] \right] \right) :+ \\
\\
+2N:Tr\left( \left[ \Phi ,Z\right] \left[ \check{\Phi},\check{Z}\right]
\right) :
\end{array}
\right] .  \notag \\
&&
\end{eqnarray}
By applying such a combinatorial formula on a generic $su(2)$-symmetric $%
\mathcal{N}=4$ SYM operator/spin-bit state $\left| S;\gamma
\right\rangle $, we obtain, after some permutational algebra and
technical tricks, the following expression for the 2-loop $su(2)$
spin-bit Hamiltonian $H_{4,sb,su(2)}\equiv H_{4,su(2)}$ :
\begin{equation}
H_{4,su(2)}=\frac{2}{N^{2}}\sum_{k,l,m=1}^{L}\left(
2P_{lm}+2P_{kl}-P_{km}-3\right) \Sigma _{klm}\left( \gamma \right) ,
\label{H4,su(2)}
\end{equation}
where $\Sigma _{klm}\left( \gamma \right) $ is the ($su(2)$)
2-loop ``twist'' operator, acting on $\zeta _{L}$, and defined as
\begin{equation}
\Sigma _{klm}\left( \gamma \right) \equiv \Sigma _{k\gamma
_{l}}\Sigma _{l\gamma _{m}}.
\end{equation}
\setcounter{equation}0
\def\theequation{7.1.\arabic{equation}}
\subsection{The planar limit}

From Eq. (\ref{sigma-decomp}), the planar limit $N\rightarrow
\infty $ affects just $\Sigma _{klm}\left( \gamma \right) $ in the
following way:
\begin{equation}
\lim_{N\rightarrow \infty }\frac{1}{N^{2}}\Sigma _{klm}\left(
\gamma \right) =\delta _{k\gamma _{l}}\delta _{l\gamma _{m}}.
\end{equation}
Therefore, we obtain
\begin{equation}
\begin{array}{l}
\lim_{N\rightarrow \infty }H_{4,su(2)}=2\sum_{k=1}^{L}\left( 4P_{k\gamma
_{k}}-P_{k\gamma _{k}^{2}}-3\right) = \\
\\
=2\sum_{k=1}^{L}\left( 4P_{k,\left[ k+1\right] }-P_{k,\left[
k+2\right] }-3\right) =H_{4,sc,su(2)}.
\end{array}
\end{equation}
Otherwise speaking, by construction the planar limit of the 2-loop
$su(2)$ spin-bit Hamiltonian coincides with the integrable $su(2)$
spin-chain Hamiltonian $H_{4,sc,su(2)}$, which in turn corresponds
to an integrable (higher-order) deformation of the previously
mentioned $XXX_{s=1/2}$ Heisenberg $su(2)$ spin-chain Hamiltonian
$H_{2,sc,su(2)}$
\cite{Beisert:2003jb,Beisert:2003tq,Beisert:2004di}.
\setcounter{equation}0
\section{$su(2)$ spin-bits beyond 2 loops}
\setcounter{equation}0
\def\theequation{8.1.\arabic{equation}}
\subsection{The ``deplanarizing operator lifts'' (d.o.l.) method}

$su(2)$-symmetric spin-chain Hamiltonians are known explicitly up
to (and including) 5 loops \cite{Beisert:2003jb,Beisert:2003tq}.
Unfortunately, combinatorial formulae for the SYM effective
vertices in the $su(2)$ sector are not known beyond 2 loops, and
therefore higher-loop $su(2)$ spin-bit Hamiltonians are not
directly obtainable as in the cases of 1 and 2 loops. Thus, other
approaches have to be pursued in order to derive them. After the
failure of the elegant and geometrically meaningful ``spin-edge
differences'' Ans\"{a}tze \cite{Bellucci:2005ma2}, the only
planarly-consistent, fully testable set of conjectures for the
higher-loop $su(2)$ spin-bit Hamiltonians are those obtained by
applying the recently proposed
\cite{Bellucci:2005ma2,Bellucci:2005ma3} ``deplanarizing operator
lifts'' (``d.o.l.") method, eventually with the additional
hypothesis of ``symmetrization of deplanarizing operator
splittings'' (hp. ``s.d.o.s.'').

In the following we will present such a deplanarizing approach in
a sketchy, algorithmic way, addressing the interested reader to
the original literature for further elucidations.

The d.o.l. algorithm may be realized through the following steps:\medskip

$1$) we have to start from the known planar results for the
Hamiltonian. Since the non-planar $1$- and $2$-loop orders are
already known and have been previously treated, we have to
consider the $3$-, $4$- and $5$-loop expressions of the\emph{\
}planar $su(2)$ spin-chain Hamiltonian \
\cite{Beisert:2003jb,Beisert:2003tq}(\emph{input }of the
deplanarizing algorithm);\medskip

$2$) then, we have to perform the (non-reductive) conventional site-index
identifications
\begin{equation}
l=k+1,\text{ }m=k+2,\text{ }...;
\end{equation}
\smallskip

$3$) therefore, we have to consider all possible products of
operators $P$'s that, in the planar limit, would give the
considered planar permutational term; this will determine some
proper ``deplanarizing operator lifts''. All such non-planar
permutational terms will come with free (real) coefficients,
constrained by two requests:\smallskip

$3.i$) their algebraic sum must give the right numerical known coefficient
of the considered planar permutational term;\smallskip

$3.ii$) they must make the spin part of the non-planar Hamiltonian
completely symmetric under the particular, inverting exchange of
site indices
\begin{equation}
\left( k,l,...,r,s\right) \leftrightarrow \left( s,r,...,l,k\right) ,
\label{exchange}
\end{equation}
as requested by the site index structure determined by the
``linking" part.

$4$) Indeed, for what concerns the ``linking" part of the
Hamiltonian, i.e. the ``twisting'' operators $\Sigma $'s, we may
generalize the ``linking" part of Eqs. (\ref{H2,su(2)}) and
(\ref{H4,su(2)}) by introducing the ($S_{L}\ni \gamma $-dependent)
``splitting and joining chain operator of order $n$'' as
\begin{equation}
\Sigma _{k_{1}k_{2}...k_{n+1}}\left( \gamma \right) \equiv \Sigma
_{k_{1}\gamma _{k_{2}}}\Sigma _{k_{2}\gamma _{k_{3}}}...\Sigma
_{k_{n-1}\gamma _{k_{n}}}\Sigma _{k_{n}\gamma _{k_{n+1}}}.
\end{equation}
\smallskip

$5$) Generally, at higher-loop orders, some (real) parameters still remain
undetermined at this step. In order to obtain a completely determined
expression of the higher-loop $su(2)$ spin-bit Hamiltonian as output of the
proposed deplanarization procedure, we may proceed as follows.

As a reasonable conjecture, we may formulate an additional
assumption, that we are going to call hypothesis of
``symmetrization of deplanarizing operator splittings'' (hp.
``s.d.o.s.''). This conjecture has to be applied \emph{after} the
symmetrization of the non-planar terms with respect to the
peculiar renaming of spin-chain site indices given by
(\ref{exchange}); it amounts to say that each of the sets of
non-planar terms arising from a considered planar term in the
deplanarization procedure will \emph{equally} contribute in the
planar limit $N\rightarrow \infty $.

For example, if, after the symmetrization with respect to (\ref{exchange}),
a planar term $\Im $ is deplanarized by the $3$-fold splitting
\begin{equation}
\Im \rightarrow a_{1}\Im _{M_{1}}+a_{2}\Im _{M_{2}}+a_{3}\Im _{M_{3}},
\label{split1}
\end{equation}
where $\Im _{M_{1}}$, $\Im _{M_{2}}$ and $\Im _{M_{3}}$ are respectively
sets consisting of $M_{1}$, $M_{2}$ and $M_{3}$ non-planar terms made by
permutation operators, then we will assume that
\begin{equation}
M_{1}a_{1}=M_{2}a_{2}=M_{3}a_{3}.
\end{equation}
Hence the contribution of $\Im _{M_{1}}$, $\Im _{M_{2}}$ and $\Im _{M_{3}}$
to the planar limit $\Im $ is the same, and therefore the operator splitting
given by (\ref{split1}) may be considered symmetric.

As it will be seen explicitly further below at 3-loops, this
additional hypothesis will allow to fix \emph{all} the free real
parameters, otherwise necessarily introduced by the deplanarizing
operator lifts acting at the considered higher-loop order. Notice
that the constraints $3.i$ and $3.ii$ are implied by the hp.
s.d.o.s. .

Thus, a complete Ansatz for the $su(2)$ spin-bit Hamiltonian at the
considered loop order is obtained\footnote{%
It should be noticed that here we assume that (eventually rather
structurally complicated) non-planar permutational terms, such
that their planar limit is zero, do \emph{not} exist; indeed, for
the time being, their existence may not be guessed by an inferring
approach starting from the planar level, such as the one adopted
in this paper.} (\emph{output }of the deplanarizing algorithm).
\setcounter{equation}0
\def\theequation{8.2.\arabic{equation}}
\subsection{Application at 3 loops}

Let us consider an explicit example of application of the
described method, in order to build a consistent Ansatz for the
expression of the $3$-loop $su(2)$ spin-bit Hamiltonian. Let us
follow the previously mentioned steps:\smallskip \medskip

$1$) we start from the known expression of $H_{6,sc,su(2)}$, namely the
3-loop, integrable, perturbative deformation of the $XXX_{s=1/2}$ Heisenberg $%
su(2)$ spin-chain Hamiltonian $H_{2,sc,su(2)}$
\cite{Beisert:2003jb,Beisert:2003tq,Beisert:2004di}, given by
(\emph{input }of the deplanarizing algorithm for $n=3$-loop order)
\begin{eqnarray}
&&
\begin{array}{l}
H_{6,sc,su(2)}= \\
\\
=4\sum_{k=1}^{L}\left[
\begin{array}{l}
15-26P_{k,k+1}+ \\
\\
+6\left( P_{k,k+1}P_{k+1,k+2}+P_{k+1,k+2}P_{k,k+1}\right) + \\
\\
+P_{k,k+1}P_{k+2,k+3}+ \\
\\
-(P_{k,k+1}P_{k+1,k+2}P_{k+2,k+3}+ \\
\\
+P_{k+2,k+3}P_{k+1,k+2}P_{k,k+1})
\end{array}
\right] ;
\end{array}
\notag \\
&&\label{D6}
\end{eqnarray}
\smallskip \medskip

$2$) we conventionally identify (without loss of generality) the spin-chain
site indices in the following way:
\begin{equation}
\text{ \ }l\equiv k+1,\text{ \ }m\equiv k+2,\text{ \ }n\equiv k+3;
\end{equation}

$3$) therefore, we have to find \emph{all} possible non-planar
permutational terms giving rise, in the planar limit $N\rightarrow
\infty $, to each of
the permutational terms of $H_{6,sc,su(2)}$ given by Eq. (\ref{D6}%
).\smallskip \medskip

$3.i$) We have that:\smallskip \medskip

$3.i.a$) the planar term $P_{k,k+1}$ receives three contributions
from the non-planar level, respectively from $P_{k,k+1}=P_{kl}$, $%
P_{k+1,k+2}=P_{lm}$ and $P_{k+2,k+3}=P_{mn}$, whence the proper
``deplanarizing operator lift'' of $P_{k\gamma _{k}}$ reads ($\xi
_{1},\xi _{2}\in R$)
\begin{equation}
-26P_{k,k+1}\rightarrow \xi _{1}P_{kl}+\xi _{2}P_{lm}-(26+\xi _{1}+\xi
_{2})P_{mn};
\end{equation}
\smallskip

$3.i.b$) the planar-level product $P_{k,k+1}P_{k+1,k+2}$ instead receives
contribution just from two non-planar terms, i.e. $%
P_{k,k+1}P_{k+1,k+2}=P_{kl}P_{lm}$ and $P_{k+1,k+2}P_{k+2,k+3}=P_{lm}P_{mn}$%
, whence the proper ``deplanarizing operator lift'' of the term $%
P_{k,k+1}P_{k+1,k+2}$ reads ($\xi _{3}\in R$)
\begin{equation}
6P_{k,k+1}P_{k+1,k+2}\rightarrow \xi _{3}P_{kl}P_{lm}+\left( 6-\xi
_{3}\right) P_{lm}P_{mn};
\end{equation}
\smallskip

$3.i.c$) analogously, for the other terms of $H_{6,sc,su(2)}$ we
obtain the following proper ``deplanarizing operator lifts'' ($\xi
_{4}\in R$):
\begin{equation}
\begin{array}{l}
6P_{k+1,k+2}P_{k,k+1}\rightarrow \xi _{4}P_{lm}P_{kl}+\left( 6-\xi
_{4}\right) P_{mn}P_{lm}, \\
\\
P_{k,k+1}P_{k+2,k+3}\rightarrow P_{kl}P_{mn}, \\
\\
P_{k,k+1}P_{k+1,k+2}P_{k+2,k+3}\rightarrow P_{kl}P_{lm}P_{mn}, \\
\\
P_{k+2,k+3}P_{k+1,k+2}P_{k,k+1}\rightarrow P_{mn}P_{lm}P_{kl};
\end{array}
\end{equation}
\smallskip

$3.ii$) thence, we impose the symmetry of the spin part under the site index
exchange
\begin{equation}
\left( k,l,m,n\right) \leftrightarrow \left( n,m,l,k\right) ;
\end{equation}
\smallskip the imposition of such a condition on the spin part decreases the
number of free (real) parameters from four to two, thence renamed $\eta _{1}$
and $\eta _{2}$;\smallskip \smallskip \medskip

$4$) finally, we put
\begin{equation}
\frac{1}{N^{3}}\Sigma _{klmn}\left( \gamma \right) =\frac{1}{N^{3}}\Sigma
_{k\gamma _{l}}\Sigma _{l\gamma _{m}}\Sigma _{m\gamma _{n}}
\end{equation}
as the linking variable part.\smallskip \smallskip

Thus, we may finally write the most general expression of the $3$-loop $%
su(2) $ spin-bit Hamiltonian ($\eta _{1},\eta _{2}\in R$):
\begin{gather}
H_{6,su(2)}\left( \eta _{1},\eta _{2}\right) =\frac{4}{N^{3}}%
\sum_{k,l,m,n=1}^{L}  \label{H6,su(2)free} \\
\notag \\
\left[
\begin{array}{l}
15+\eta _{1}\left( P_{kl}+P_{mn}\right) -2\left( \eta _{1}+13\right) P_{lm}+
\\
\\
+\eta _{2}\left( P_{kl}P_{lm}+P_{mn}P_{lm}\right) + \\
\\
+\left( 6-\eta _{2}\right) \left( P_{lm}P_{kl}+P_{lm}P_{mn}\right) + \\
\\
+P_{kl}P_{mn}-P_{kl}P_{lm}P_{mn}-P_{mn}P_{lm}P_{kl}
\end{array}
\right] \Sigma _{k\gamma _{l}}\Sigma _{l\gamma _{m}}\Sigma _{m\gamma _{n}}.
\notag
\end{gather}

Formulating the hp. s.d.o.s.\emph{\ }we get $\left( \eta _{1},\eta
_{2}\right) =\left( -\frac{13}{2},3\right) $, and therefore we obtain a
\emph{completely fixed} expression for the $3$-loop $su(2)$ spin-bit
Hamiltonian (\emph{output }of the deplanarizing algorithm for $n=3$-loop
order):
\begin{gather}
H_{6,su(2)}=\frac{4}{N^{3}}\sum_{k,l,m,n=1}^{L}  \label{H6,su(2)} \\
\notag \\
\left[
\begin{array}{l}
15-\frac{13}{2}\left( P_{kl}+P_{lm}+P_{mn}\right) + \\
\\
+3\left( P_{kl}P_{lm}+P_{mn}P_{lm}+\right.  \\
\\
\left. +P_{lm}P_{kl}+P_{lm}P_{mn}\right) + \\
\\
+P_{kl}P_{mn}-P_{kl}P_{lm}P_{mn}-P_{mn}P_{lm}P_{kl}
\end{array}
\right] \Sigma _{k\gamma _{l}}\Sigma _{l\gamma _{m}}\Sigma _{m\gamma _{n}}.
\notag
\end{gather}

Analogous, more and more involved, expressions for the $4$- and
$5$-loops $su(2)$ spin-bit Hamiltonian (with or without the
additional hp. s.d.o.s.) have been obtained by applying the d.o.l.
method \cite{Bellucci:2005ma2,Bellucci:2005ma3}.
\setcounter{equation}0
\def\theequation{\arabic{section}.\arabic{equation}}
\section{Outlook and further developments}

The d.o.l. method \cite{Bellucci:2005ma3} is fully compatible with
(independently obtained) known results at the $1$- and $2$-loop,
non-planar level
\cite{Bellucci:2004ru,Bellucci:2004qx,Bellucci:2005ma1}, and it
allows one to obtain explicit formulae for the $3$-, $4$- and $5$-loop $%
su(2) $ spin-bit Hamiltonians. By construction, such expressions
are planarly-consistent, i.e. they have the correct planar limit,
matching the known results reported in the literature (see e.g.
\cite {Beisert:2003jb,Beisert:2003tq,Beisert:2004di}).\smallskip

It is also worth noticing that, by construction, all the
higher-loop $su(2)$ spin-bit Hamiltonians (for 3-loops, see Eqs.
(\ref{H6,su(2)free}) and (\ref {H6,su(2)})) show an explicit full
factorization in the spin and chain-splitting parts; as already
pointed out in \cite{Bellucci:2005ma1}, such a property is
expected to hold at every loop order, since the Hilbert space of
the spin-bit model $\mathcal{H}_{sb}$ is given by the direct
product (modulo the action
of the permutation group $S_{L}$) of the spin-chain Hilbert space $\mathcal{H%
}_{sc}$ and of the linking space $\zeta _{L}$.\smallskip

Attention must also be paid to the fact that, while (both at
non-planar and planar level) the $1$- and $2$-loop formulae for
the $su(2)$ Hamiltonians are \emph{linear} in the site permutation
operators $P$'s, the $3$-, $4$- and $5$-loop level expressions,
both at non-planar and planar level, show a non-linearity (and
non-linearizability) in $P$'s. For example, the
non-linearizability of the $3$-loop $su(2)$ spin-chain Hamiltonian (%
\ref{D6}) caused the failure of the elegant and geometrically
meaningful ``spin edge-differences'' \cite{Bellucci:2005ma2}
approach to higher-loop Ans\"{a}tze.

Thus, the non-linearity (and non-linearizability) in site permutation
operators seems to be a crucial and fundamental feature, starting to hold at
the $3$-loop order, of the spin part of the Hamiltonian of the $su(2)$ spin-bit model, underlying the non-planar dynamics of the $su(2)$ sector of the $\mathcal{N}%
=4,d=4$ SYM theory. Reasonably, one would expect that such a
breakdown of ``permutational linearizability'' at $3$ loops (for
the first evidences from $3$-loop calculations, see e.g. \cite
{Beisert:2003jb,Beisert:2003tq}; for further subsequent
developments see e.g. \cite
{Tseytlin:2004xa,Tseytlin:2003ac,D'Hoker:2003}%
) gives rise, by means of the AdS/CFT correspondence \cite
{Maldacena:1998re,Gubser:1998bc}, to some ``new'' features in the
dynamics of the (closed) superstrings in the bulk of
$AdS_{5}\times S^{5}$. Actually, in the AdS/CFT correspondence
framework, there is a problem of discrepancy between the
calculations made with fast spinning, semiclassical strings (i.e.
in the so-called Frolov-Tseytlin limit) and the calculations made
in the thermodynamical limit of long spin-chains with a large
number of excitations (i.e. the so-called
Berenstein-Maldacena-Nastase limit) (see e.g.
\cite{Beisert:2004di} and Refs. therein). Such a disagreement
starts to hold at 3 loops, and it is one of the most intriguing
``mysteries" of the AdS/CFT conjecture. Recently, an explanation
for such $3$- and higher-loop disagreement has been proposed: it
should be related to an ``order-of-limit" non-commutation problem
in the perturbative expansions and thermodynamical asymptotical
regimes or, equivalently, to the presence of operational
``wrapping" interactions (see e.g.
\cite{Beisert:2004di,Beisert:2004cr}). Conjecturally, we may here
put forward the suggestion that the breakdown of ``permutational
linearizability'' of $su(2)$ spin-chain/spin-bit Hamiltonians,
which starts to hold at $3$ loops, could be related to such a
``$3$-loop discrepancy mystery" in AdS/CFT, and possibly it could
be extended also to larger symmetries inside $psu(2,2|4)$.
\smallskip

Also, the application of the d.o.l. method, originally introduced
for the Hamiltonian, to the higher-order charges of the $su(2)$
spin-chain model \cite{Bellucci:2005ma3} raises some interesting
and intriguing questions, such as:\smallskip

$1$) the d.o.l. appears to be consistent only for odd higher-order
charges, sharing the symmetry properties of the Hamiltonian. Thus,
the extension of the deplanarization procedure to even
higher-order charges, and in general to antisymmetric operators,
should be needed, in order to have a complete deplanarizing
algorithm for the local operators in $\mathcal{N}=4,d=4$ SYM
theory;\smallskip

$2$) we know that the spectrum of the (perturbatively expanded)
Hamiltonian of the spin-chain/spin-bit model is related to the
spectrum of the (perturbatively expanded) anomalous dimensions and
mixing of local operators on the $\mathcal{N}=4,d=4$ $SU(N)$ SYM
gauge theory side of AdS/CFT. Then a natural question
\cite{Beisert:2003tq} to ask is: \emph{do the spectra of the
higher-order planar charges (and of their deplanarized
counterparts) have a physical meaning in }$\mathcal{N}=4$\emph{\
SYM ?}

The deep question is evidently: \emph{why does exact integrability
seem to hold for every loop-order in the planar
}$\mathcal{N}=4,d=4$\emph{\ SYM} \cite{Serban:2004jf,Beisert:2004di,Aru,Beisert:2005st}\emph{, and why and how is it lost at the non-planar level?%
}

By the way, even if the exact (classical) integrability is lost
when deplanarizing, i.e. when passing from the (all-loop) $su(2)$
Heisenberg spin-chain to the (all-loop) $su(2)$ spin-bit model,
nevertheless we may put forward the following intriguing
suggestion: could the $su(2)$ spin-bit model still be an
integrable model, but in a sort of generalized,
broader sense, e.g. in the sense of \emph{quasi-integrability} and \emph{%
quasi-exact solvability} (see e.g. \cite
{Falqui:1992}) or \emph{%
quantum-integrability} (see e.g. \cite {Kazakov:2004})?

If yes, then in general the deplanarization procedure here
presented might algebraically correspond to some kind of
\emph{``deformation''} of the (dynamical) symmetries of the system
being considered, and thence of the structure and properties of
its (eventually conserved) charges.

In our opinion, this is an interesting problem, strictly related to the
consistent definition of the higher-order charges of the spin-bits as the
deplanarization of the infinitely many conserved charges of the spin-chains,
and it is currently under study.

Moreover, we notice that it would be interesting, following recent
research directions, to extend the considered deplanarizing method
to other operatorial sectors of the $\mathcal{N}=4,d=4$ SYM theory
\cite {Bellucci:2005np1}. Indeed, sectors with non-compact
symmetries have been shown to be relevant, in order to describe
the renormalization in the large $N_{c}$ (non-)SUSY QCD, also in
relation with the attempts to construct a string description of
QCD (see e.g. \cite {Ferretti:2004-1} and Refs. therein).

Finally, some possible additional directions for further research
are briefly summarized as follows:

- All the spin-chain/spin-bit models treated so far are
characterized by periodic boundary conditions, corresponding to
closed chains. The possibility to modify the boundary conditions
yields open spin-chain models, corresponding to a dynamical
discretization of the open strings \cite {Wu:2004jh}. The issue of
deplanarizability, and the related problem of integrability, of
such models remains to be discussed.

- In general, the spin-bits may be considered as a dynamical
polymer model with decaying and fusing chains; potential
applications to relativity theory \cite{Ashtekar:1996int}, field
theory \cite{Bergman:1997npb} and biophysics could be addressed.

- Interesting analogies could be explored with the spin-network
approach to discrete quantum gravity \cite{Penrose:1971}.

- The above mentioned non-planar permutational identities could be linked to
the random graph theory on a lattice \cite{Bellucci:2005ma2}.

- The spin representation of the permutation operators in the
$su(2)$ sector leads to some ``generalized'' Fierz identities for
Pauli $\sigma $ matrices \cite{Bellucci:2005ma2}, deserving a more
detailed analysis.

- An alternative approach to the calculation of anomalous dimensions in $%
\mathcal{N}=4$ SYM is based on matrix models \cite {Agarwal}, which
can be also formulated in terms of a non-commutative field theory on
a torus \cite {Bellucci:2004rua,Sochichiu:2002}, and whose
equivalence with
spin-chain/spin-bit models in the $N\rightarrow \infty $ (planar) and $%
N\rightarrow 0^{+}$ limits is still under study. An interesting
direction of research to be pursued would be the formulation of
such matrix models on other ``fuzzy" manifolds, such as the
``fuzzy sphere" \cite {Valtancoli:2002}, and the study of the
relation of such non-toroidal ``fuzzy" matrix models with the
initial non-commutative torus representation.

\subsection*{Acknowledgements}

Most of the results here presented have been obtained during the
last years by the LNF team, directed by S. Bellucci, and composed,
beside the author, by P.-Y. Casteill, F. Morales and C. Sochichiu.
I would like to gratefully acknowledge stimulating discussions
with all of them.

\end{document}